\documentclass[prb,aps,twocolumn,superscriptaddress,showpacs]{revtex4-2}

\usepackage[T1]{fontenc}
\usepackage{ae}   
\usepackage{wrapfig}
\usepackage{float}
\usepackage{amsmath}
\usepackage{amssymb}
\usepackage{amsfonts}
\usepackage{mathtools} 
\usepackage[pdftex]{graphicx}	
\usepackage{array}	
\usepackage{bm}		
\usepackage{dsfont}	
\usepackage{ulem}
\usepackage{color}
\usepackage{hyperref}
\usepackage{units}
\usepackage{gensymb}
		
\newcommand{\br}{{\bf r}}

\begin{document}

\title{Friedel Oscillations and superconducting-gap enhancement by impurity scattering}
\author{Matthias Stosiek}
\affiliation{Physics Division, Sophia University, Chiyoda-ku, Tokyo 102-8554, Japan}
\author{Clemens Baretzky}
\affiliation{Physikalisches Institut, Karlsruhe Institute of Technology, Wolfgang-Gaede Str. 1, 76131 Karlsruhe, Germany}
\affiliation{Freiburg Materials Research Center (FMF), University of Freiburg, Stefan-Meier-Str. 21, 79104 Freiburg, Germany}
\author{Timofey Balashov}
\affiliation{Physikalisches Institut, Karlsruhe Institute of Technology, Wolfgang-Gaede Str. 1, 76131 Karlsruhe, Germany}
\author{Ferdinand Evers}
\affiliation{Institute of Theoretical Physics, University of Regensburg, 93040 Regensburg, Germany}
\author{Wulf Wulfhekel}
\affiliation{Physikalisches Institut, Karlsruhe Institute of Technology, Wolfgang-Gaede Str. 1, 76131 Karlsruhe, Germany}
\date{\today }

\begin{abstract}
Experiments observe an enhanced superconducting gap over impurities as compared to the clean-bulk value. In order to shed more light on this phenomenon, we perform simulations within the framework of Bogoliubov-deGennes theory applied to the attractive Hubbard model. 
The simulations qualitatively reproduce the experimentally observed enhancement effect; it can be traced back to an increased particle density in the metal close to the impurity site. 
In addition, the simulations display significant differences between a thin (2D) and a very thick (3D) film. In 2D pronounced Friedel oscillations can be observed, which decay much faster in (3D) and therefore are more difficult to resolve. Also this feature is in qualitative agreement with the experiment. 
\end{abstract}


\maketitle

\section{Introduction}

A non-magnetic charged impurity in a metal at low temperatures is screened by Friedel oscillations in the particle density \cite{friedel52}. A related effect was predicted more than five decades ago in conventional superconductors in the presence of a non-magnetic impurity, where the superconducting gap shows an oscillatory response \cite{fetter65}. The cause of both phenomena is the presence of a sharp Fermi surface in these systems.
The description of such a response in the superconducting system is considerably more complex than in the non-interacting case. This is due to the self-consistency requirement, as it arises in the mean-field description of an interacting system \cite{tanaka02}.
When translational symmetry is broken, the non-linear nature of mean-field Hamiltonians  severely limits analytical approaches and makes numerical simulations computationally demanding.

While analytical progress has been made \cite{fetter65, machida71, flatte96, balatsky06, torre16, schmalian18}, the analytical form of the response of the superconducting gap is known only in the case of an impurity in the bulk of a 3D system \cite{fetter65, flatte96}.
Numerical studies \cite{flatte96, tanaka2000, tanaka02, tsai09, schmalian18} on the other hand were limited by small system sizes and key properties, such as the spatial decay of the response, have not been analyzed quantitatively. 

To map local variations of the density of states due to scattering, i.e. Friedel oscillations, scanning tunneling microscopy (STM) in ultra high vacuum (UHV) is the ideal approach \cite{Hasegawa1993}. At low enough temperatures, it also allows determination of the local superconducting gap $\Delta$. Local scatterers were created by short Ar ion sputtering of a clean bulk Al(111) sample. Ions hitting the surface produce defects, mostly implanted Ar \cite{Schmid2001} acting as pure potential scatterers without spin.
At the base temperature of the STM of $\approx$ 25 mK \cite{Balashov18}, spectra of the voltage dependent differential conductance $dI/dU$ were recorded using non-superconductive W tips. 

\begin{figure}[t]
\centering
\includegraphics[width=1.0\columnwidth]{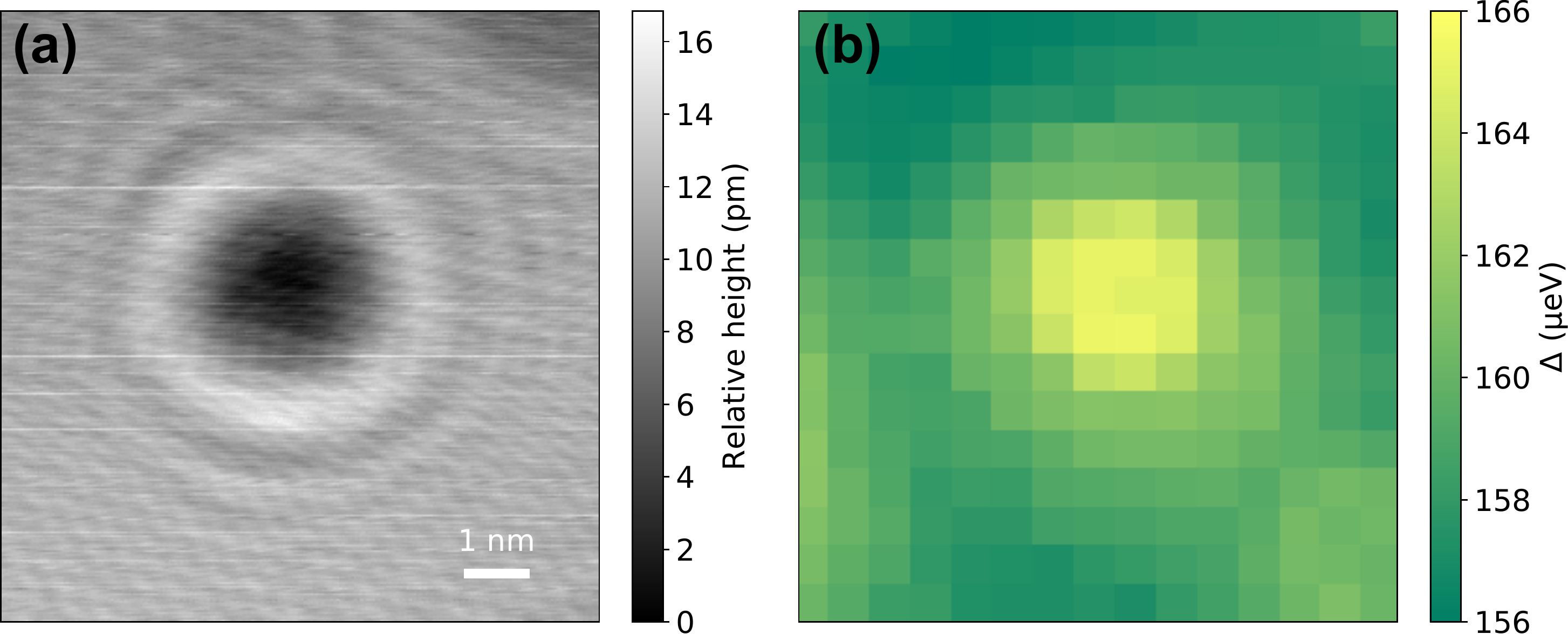}
\caption{(a) STM topography of an Al(111) surface with a buried defect after mild Ar$^+$ bombardment. (b) Superconducting gap $\Delta$ obtained by fitting $dI/dV$ spectra recorded as function of lateral position; further experimental details are given in App. \ref{appendix_experiment}. 
}
\label{Fig3}
\end{figure}

The experimental situation concerning the Friedel oscillations around an implanted Ar impurity is illustrated in Figure \ref{Fig3} a). Clearly, Friedel oscillations are visible around the defects \cite{Schmid2001}. 
Figure \ref{Fig3} b) displays a 2D-map of $\Delta(x,y)$ near a  defect. Interestingly, the superconducting gap is enhanced near the impurity site.  A similar gap-enhancement is also observed in measurements adopting Fe-adatom as impurities, see Appendix \ref{appendix_experiment}.

Concerning numerical studies of gaps near impurity sites, the published numerical data exhibits local decreases \cite{flatte96,tsai09,schmalian18} and increases \cite{tanaka2000, tanaka02} relative to the unperturbed gap; the systematics has not yet been analyzed. 

Motivated by the measurements shown in Fig. \ref{Fig3}, we have performed numerical simulations of Friedel oscillations in thin (2D) and thick (3D) superconducting films in unprecedented system sizes. We work with a tight-binding model parametrized for an adatom on an Aluminum substrate, 
adopting s-wave superconductivity on the Bogoliubov-deGennes level.
For comparison, earlier studies did not consider adatoms but substitutional atoms in the bulk or near surfaces. Our choice is motivated by the typical arrangement in STM-measurements.

Our most important conclusions are two-fold: First, the gap-enhancement may be understood as a consequence of the particle density underneath the impurity being increased as compared to the unperturbed surface. Second, depending on the film thickness Friedel oscillations exhibit a qualitatively different behavior with a $1/r$-envelope for 2D and a decay considerably faster than $1/r^2$ in 3D. 
Our findings thus explain the most striking features seen in experiment, Fig. \ref{Fig3}. 

\section{Theory}

\subparagraph*{Model and method.}
We study the Hubbard model with attractive interaction $U$ dressed with an Anderson impurity realized as an extra site ("adatom") with 
scattering potential $V_\text{imp}$. 
The model is solved on square (2D) and cubic (3D)  lattices on mean-field level, i.e., within the Bogolibov-deGennes (BdG) approximation treated with full self-consistency and stipulating $s$-wave pairing.  
The interaction strength $U$ is tuned so that the superconducting correlation length equals half the system size, $\xi=54a$ (2D) and $\xi=10.6a$ (3D), where $a$ denotes the lattice spacing.
Further computational details have been relegated to the 
Appendix \ref{a2}.   

\begin{figure}[t]
\includegraphics[width=0.47\linewidth]{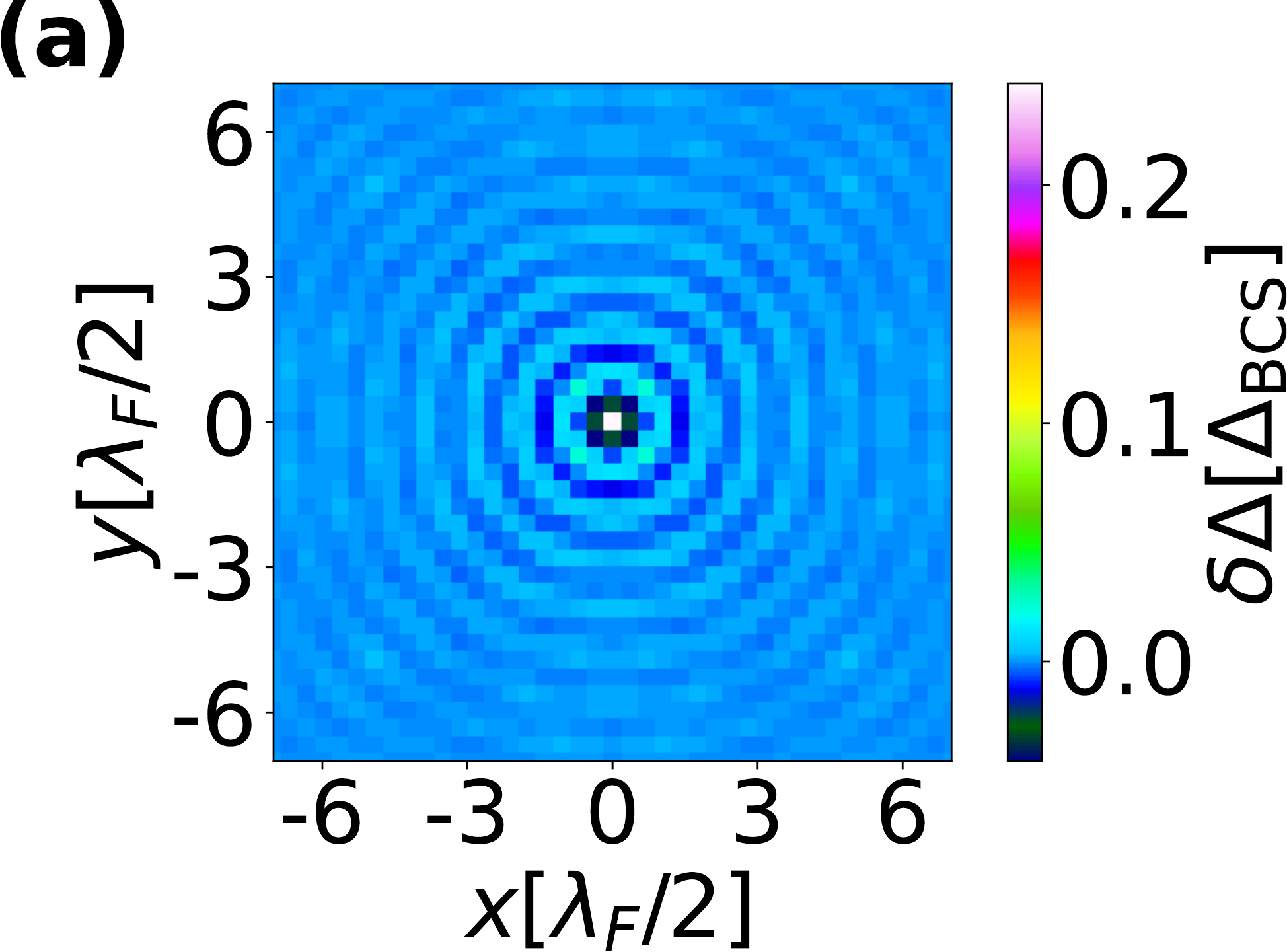}
\hspace{0.1cm}
\includegraphics[width=0.49\linewidth]{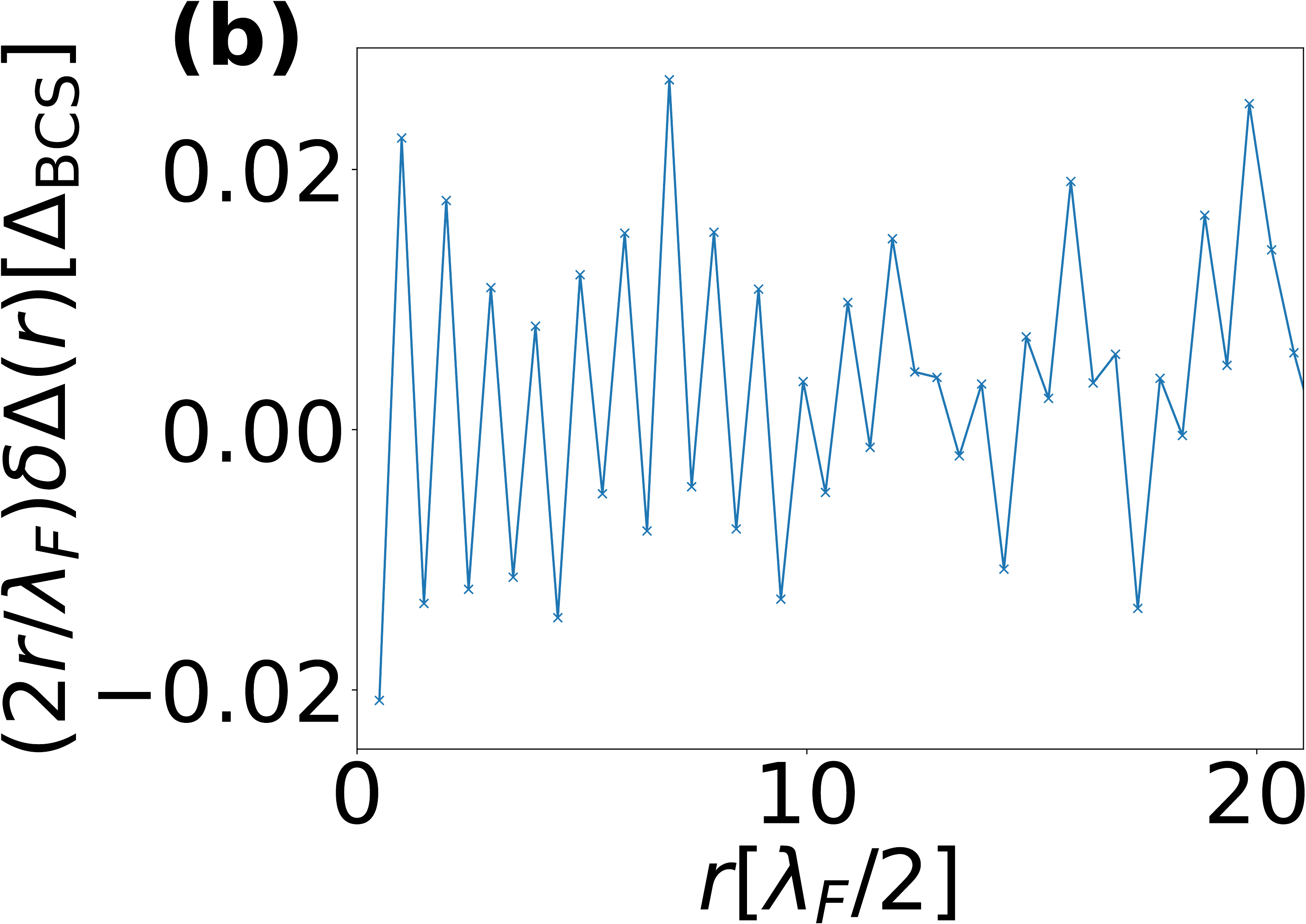}
\caption{(a) Spatial map of the response of the pairing amplitude of a 2D superconductor to an Anderson impurity located at the origin.  
 (b) Response of the pairing amplitude along the surface diagonal multiplied by the distance from the impurity $r$ in units of the lattice constant $a$.
  (Parameters: filling $n=0.2$, $U=1.6$, $V_\text{imp}=-0.06t$, $\lambda_F/2\approx 2.85a$) }
\label{Fig4}
\end{figure}

\paragraph{\bf Results -- Thin-film superconductors (2D).}
Fig. \ref{Fig4} (a) displays the spatially resolved response of the pairing amplitude $\Delta(\br)$ near the Anderson impurity. 
Friedel oscillations with frequency of $2k_F$ are clearly visible, as one would have expected. Also, the superconducting gap is enhanced at the impurity site, in the computation by more than $24\%$.  

In Fig. \ref{Fig4} (b) the envelope of the oscillatory part of the pairing amplitude is analyzed. In order to highlight the $1/r$-type power-law decay, the product $r\Delta(r)$ is plotted; it displays amplitude fluctuations with a strength independent of $r$ as characteristic of a $1/r$ envelope. 
Note that this behavior is in pronounced contrast to the familiar textbook result for the free Fermi gas, where we have the asymptotics $\sin(2k_{\rm F}r)/(k_{\rm F}r)^d$ for particle-density oscillations in $d$ dimensions, so $1/r^2$ in 2D \cite{gabrielebook}. 
We mention that a similar observation can be made in layered superconductors with anisotropic pairing, where the RKKY interaction is comprised of a normal $r^{-2}$ and a superconducting $r^{-1}$ contribution \cite{aristov97}.

\paragraph{\bf Results -- Thick-film superconductors (3D).}

\begin{figure}[t]
\includegraphics[width=0.45\linewidth]{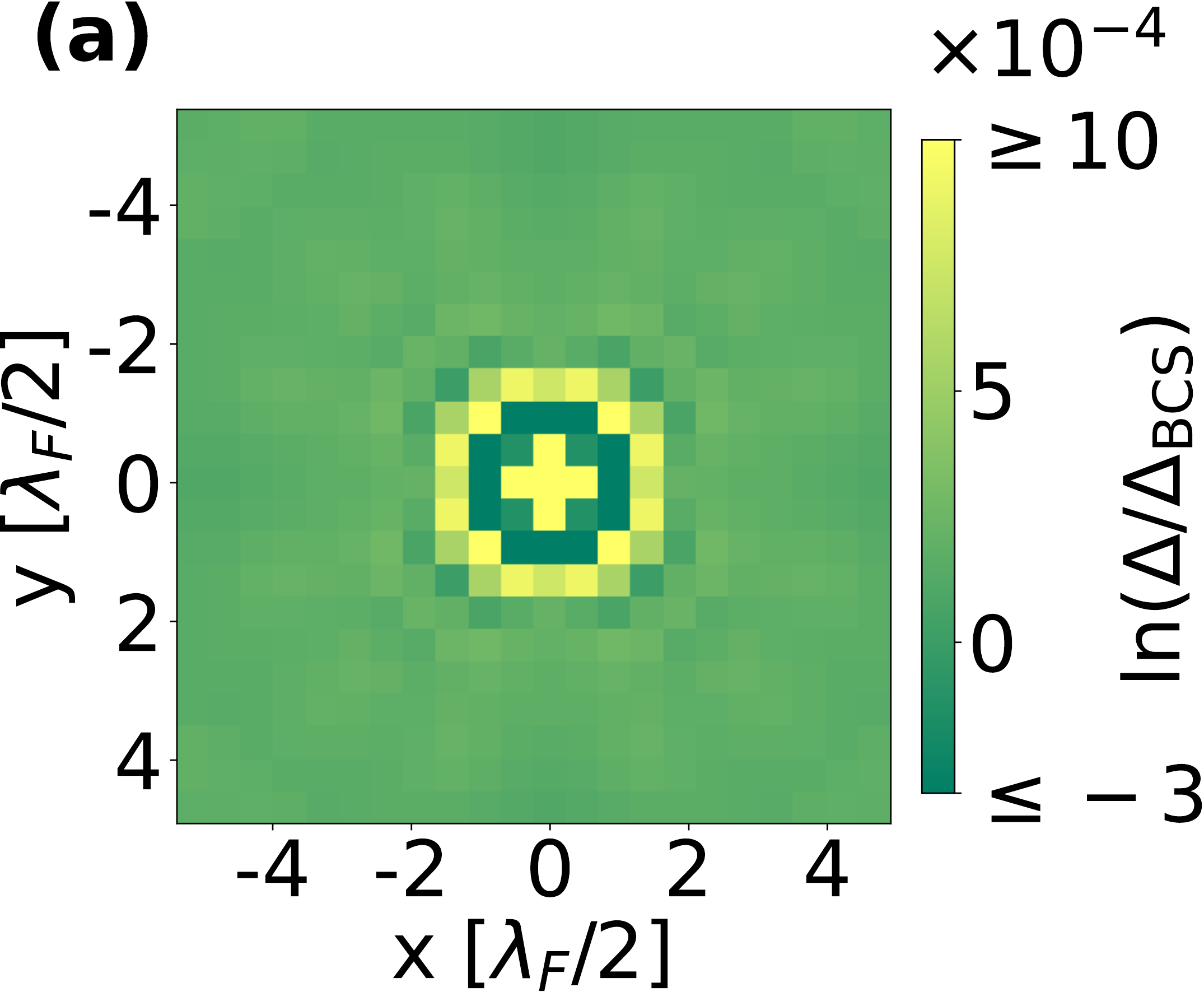}
\includegraphics[width=0.53\linewidth]{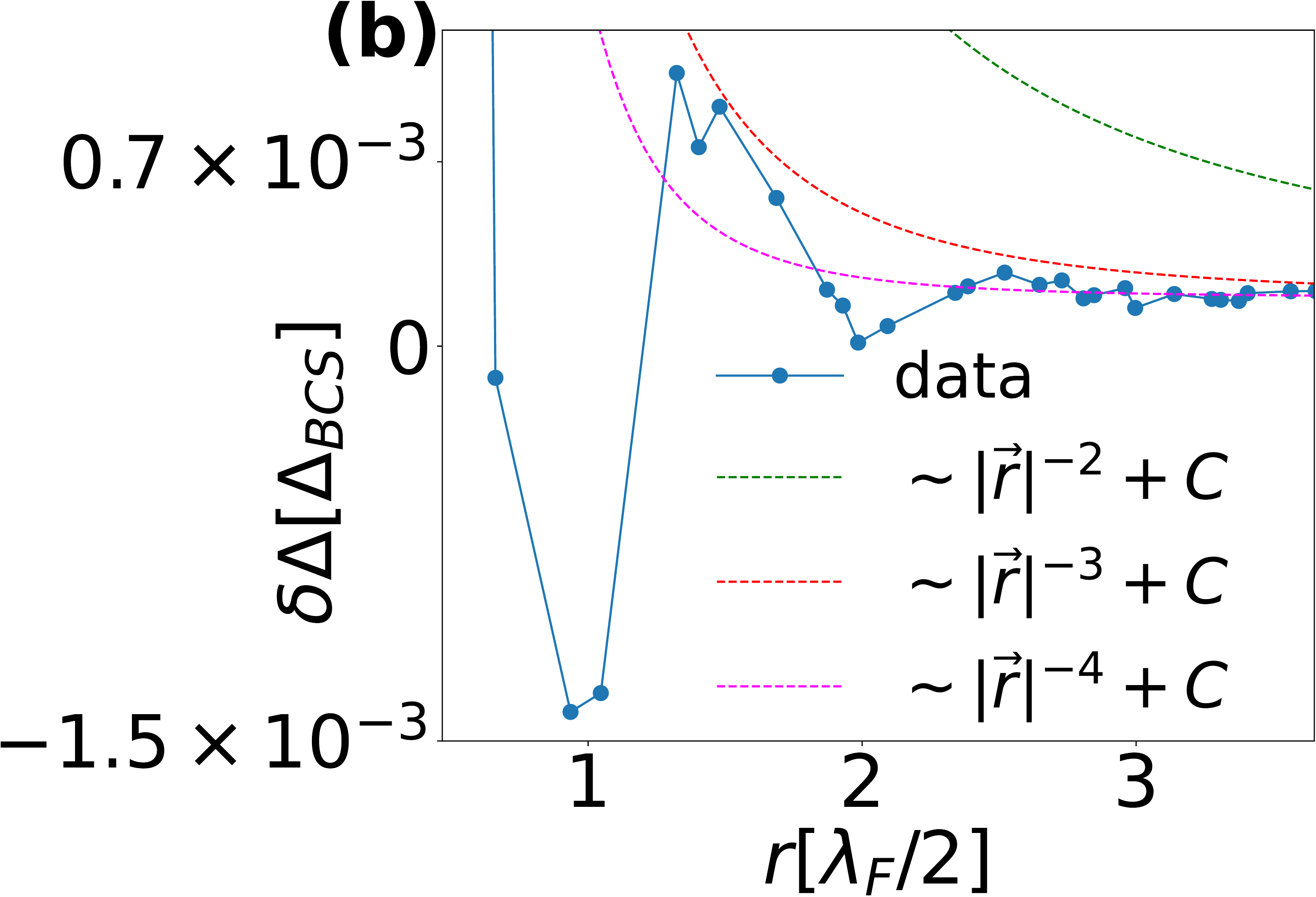}
\caption{(a) Spatial map of the logarithm of the pairing amplitude in the surface layer of a 3D superconductor with an Anderson impurity at the origin. (b) Response of the pairing amplitude to an impurity on the surface of a 3D superconductor. The response is computed from data shown in Fig. \ref{Fig6} (a) averaged over the surface angle. The prefactors of the power-law fits are chosen such that they agree with the data at $r=1a\approx 0.47\lambda_\text{F}/2$.
(Parameters: filling $n=0.12$, $U=3.2$, $V_\text{imp}=-0.06t$, $\lambda_F/2 \approx 2.14a$, $C\approx 1.8 \times 10^{-4} \Delta_\text{BCS}$)
}
\label{Fig6}
\end{figure}
 In Fig. \ref{Fig6} the response of the pairing amplitude to an impurity situated in the center of the surface layer of a cube with linear size $L{=}22$ sites is shown. The overall shape of the response is qualitatively similar to the 2D case. For instance, an enhancement of the gap by $20\%$ is observed underneath the impurity site, $\Delta(r{=}0){=}1.2\Delta_{\rm BCS}$.
 
 A  significant difference with the thin-film response occurs with respect to the decay of the oscillations. In order to better highlight this important point, in Fig. \ref{Fig6}(b) we plot the response of the local gap function, $\delta \Delta$, averaged over a circle with radius $r$ around the impurity site. Due to the cubic symmetry underlying our microscopic model (and the simulation volume), the rotational symmetry is broken and therefore the integration over the circle reduces the oscillation amplitudes. Nevertheless, the $2k_F$ oscillation is clearly visible in Fig. \ref{Fig6}(b). 
As to the corresponding envelope function, we observe a significantly different thick-film decay  compared to the thin-film limit. While for the system sizes available to us the true asymptotics is out of reach, in the intermediate regime our data in Fig. \ref{Fig6} exhibits an envelope $r^{-3}+C$; the constant $C$ depends on system size and is expected to vanish upon $L\to\infty$, ultimately revealing the true asymptotic decay.

\begin{figure}[t]
\includegraphics[width=0.48\linewidth]{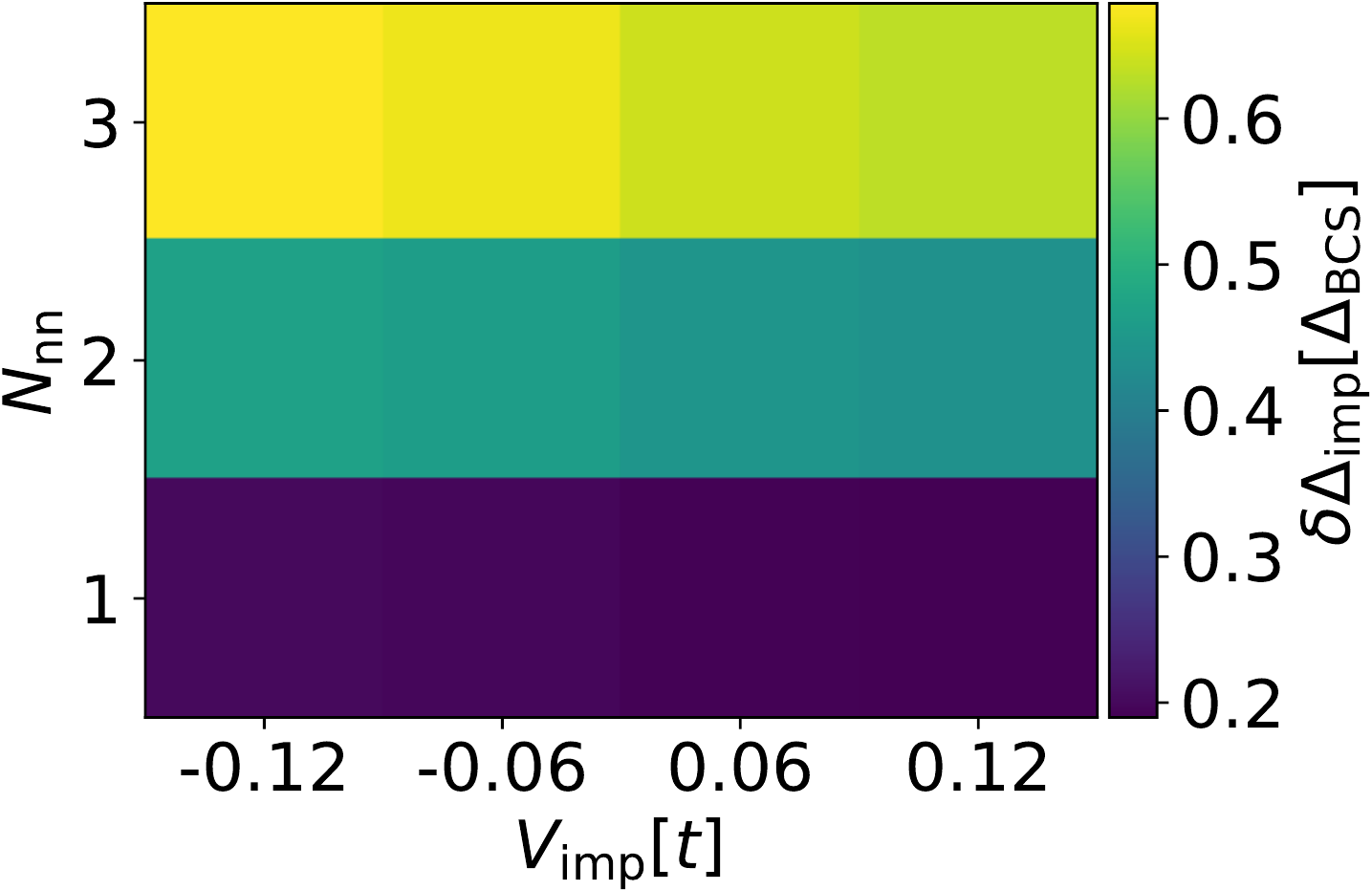}
\includegraphics[width=0.48\linewidth]{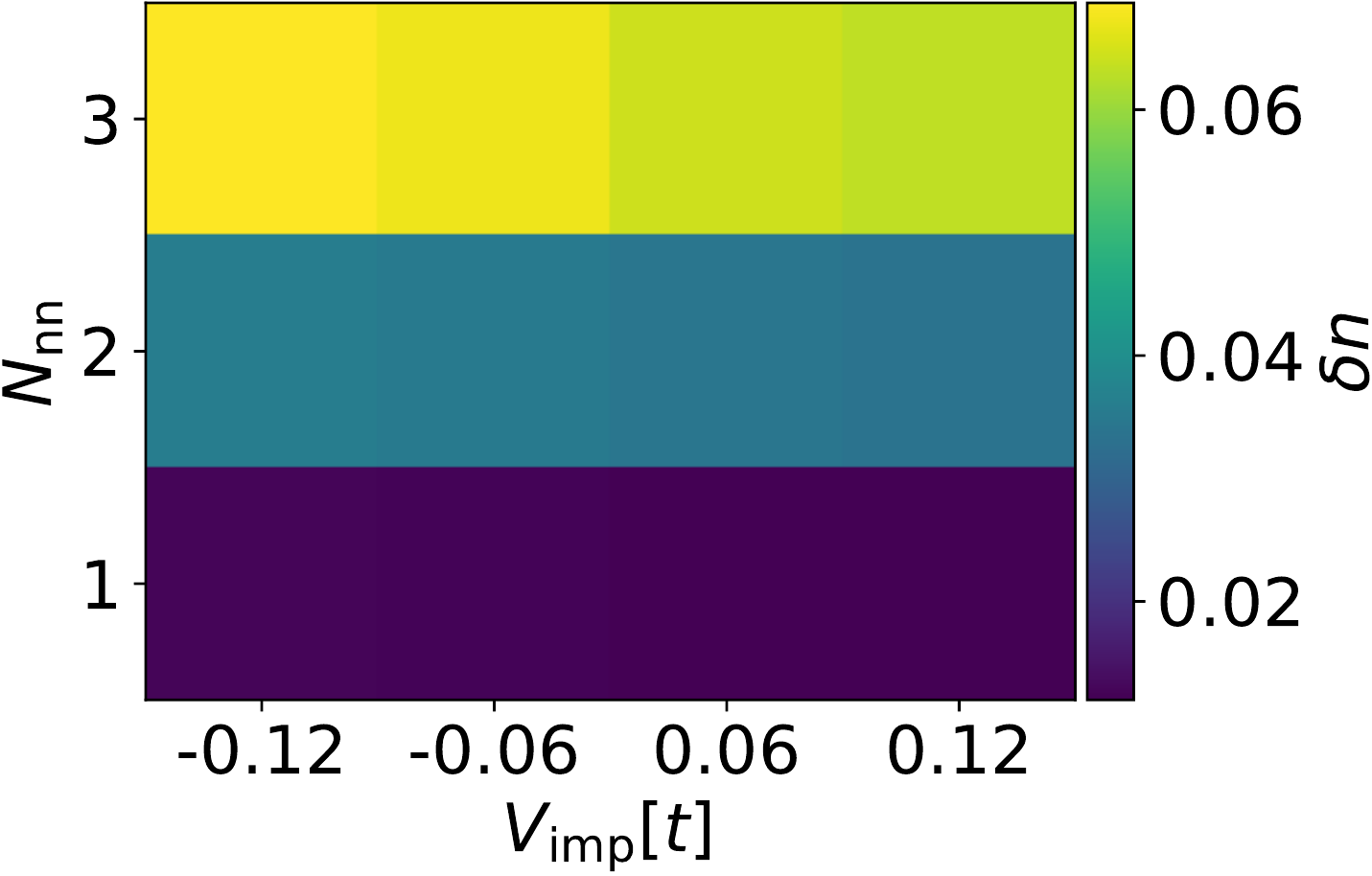}
\caption{Response of the superconducting gap,  $\delta\Delta_\text{imp}$ (left), and the particle density, $\delta n$ (right), underneath the impurity (adatom) located on the surface of the cubic lattice (same situation as in Fig. \ref{Fig6}). Shown is the dependency on the potential $V_\text{imp}$ (horizontal axis) and the number of nearest neighbors $N_\text{nn}$ of the impurity site (vertical axes).  $\delta\Delta_\text{imp}$ is the average of the superconducting gap on all sites neighboring the impurity. 
(Parameters: filling $n=0.12$, $U=3.2$)
}
\label{Fig5}
\end{figure}

\subparagraph*{Gap and density.} We rationalize our results by recalling a basic result from BCS theory that connects the superconducting gap with the density of states at the Fermi surface $\rho_\text{F}$: 
\begin{align}
    \Delta\simeq 2\hbar \omega_\text{D}  e^{-\frac{1}{U\rho_\text{F} a^d}}. 
    \label{eq1} 
\end{align}
Here, the pairing volume $a^d$ relates the attractive Hubbard interaction $U$ to the effective pairing  interaction $V_\mathbf{k,k'}$ featured by BCS theory. The latter binds a Cooper pair with energy density $V$ if the corresponding particle-hole states are situated within a shell around $E_\text{F}$ of width  $\hbar\omega_\text{D}$ (in the present model $\hbar\omega_\text{D}\approx U$); hence: $a=v_\text{F}/
\omega_\text{D}$ and $V\sim Ua^d$, introducing the Fermi-velocity $v_\text{F}$. Reinterpreting Eq. \eqref{eq1} on a local scale, we stipulate an approximate relation  
\begin{align}
    \ln\left[ \frac{\Delta(\mathbf{r})}{2\hbar \omega_\text{D}}\right]\simeq    -\frac{1}{N(\mathbf{r})}, 
    \label{e1} 
\end{align}
where the abbreviation $N(\mathbf{r})\coloneqq U\rho_\text{F}(\mathbf{r}) a^d$ has the interpretation of the number of particles within the distance $\sim U$ from $E_\text{F}$ 
inside the correlation volume $a^d$.  In the superconducting phase these particles participate in pair-formation with a gap-size exponentially decreasing with $N^{-1}(\mathbf{r})$ increasing. These heuristic considerations prompt the formulation of a rule of thumb: a local modulation of the particle density in the normal phase is typically accompanied by a proportional modulation of the local gap function in the superconducting phase; in other words, particle densities $n(\br)$ slightly enhanced above the clean reference value correspond to slightly enhanced gap-values $\Delta(\br)$. 

A numerical test of this propositon has been depicted in Fig. \ref{Fig5}. For different kinds of binding scenarios - adatoms in on-site, bridge and hollow position ($N_\text{nn}=1,2,3$) - and for varying on-site potential $V_\text{imp}$, the response of the density and gap in the metal substrate near the impurity is shown. The conjectured rule of thumb manifests in the strong resemblance of left and right panels.  

\begin{figure}[t]
\centering 
\includegraphics[width=0.9\columnwidth]{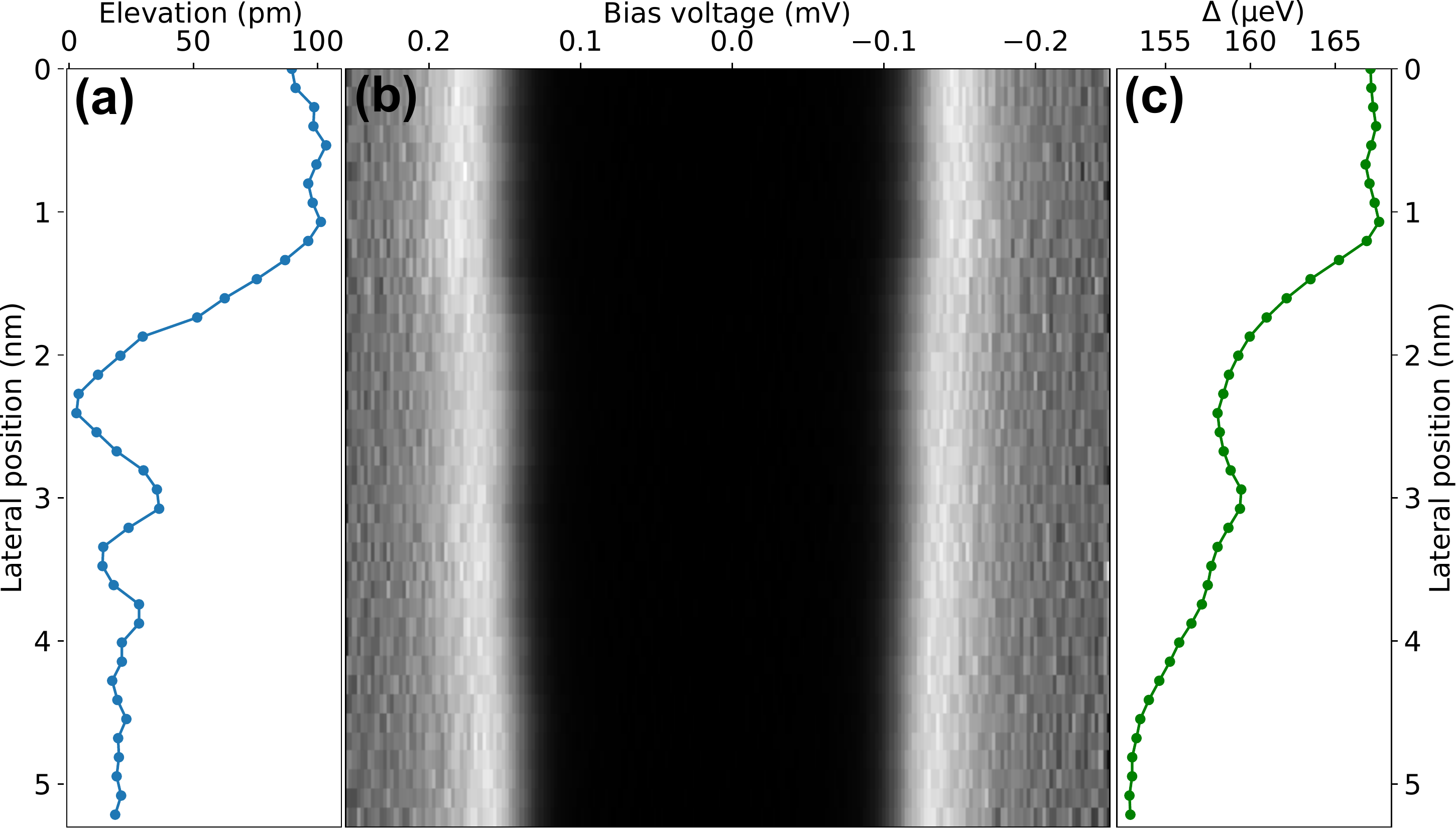}
\caption{(a) Change of the z-position, i.e. tip elevation, with distance to the surface defect. (b) Greyscale encoded $dI/dV$ recorded as function of distance from the impurity on the same lateral positions as a). (c) Superconducting gap $\Delta$ fitted to data of b) as function of position from the impurity.    
}
\label{Fig2} 
\end{figure}

\subparagraph{Further discussion.} 
Our simulation results on the 3D-Friedel oscillation are qualitatively reproducing aspects of our measurements. Figure \ref{Fig2}a) shows the z-position of the STM tip as function of distance from the center of a defect that shows up as a protrusion. In the topography, the Friedel oscillations are again found followed at larger distances by a fall back of the elevation to that of the plane surface. Their period of the oscillations is with $\approx$0.9 nm in good agreement with the literature \cite{Sferco2007,Shiihara2010}. Fig. \ref{Fig2}b) displays $dI/dU$ spectra obtained as function of distance from the defect encoded in greyscale. Black corresponds to vanishing $dI/dU$ inside the gap, while the coherence peak appears bright. Clearly, the gap is significantly enhanced atop of the defect. Upon moving away from the defect, the gap decreases with slight oscillations and reaches the clean value some 5 nm away from the defect as evidenced by the fitted values of $\Delta$ in Fig. \ref{Fig2}c). Note that the variations are laterally much finer than the coherence length of Al of 1.6 $\mu$m at this temperature \cite{Kittel} and happen on the length scale of the variations in the density of states.      

\section{Conclusions and Summary}
Motivated by STM-measurements of the superconducting gap near impurities on an Al(111) surface, we have performed corresponding simulations within the BdG-formalism. Our simulations capture all qualitative features seen in the experiment: (i) relatively quickly decaying Friedel oscillations (as compared to the thin-film limit studied in simulations); (ii) an enhancement of the superconducting gap in the vicinity of the impurity of the order of 10\%. The enhancement has been traced back to the modulation of the (normal state) particle density in the vicinity of the impurity. 

Our results are encouraging in the sense that BdG has been shown to provide a reliable framework reproducing the salient features of physical reality, at least on a semi-quantitative level. Future work should provide more detailed (and extensive) investigations comprising, in particular, a larger set of experimental realizations screening different substrates and adatoms. 

\subparagraph*{Acknowledgments.}
We are grateful to Igor Burmistrov and Christoph Strunk for numerous inspiring discussions and Thomas Gozlinsky for preparing some of the figures.
We acknowledge funding from the Deutsche Forschungsgemeinschaft (DFG) with grants Wu 394/12-1, INST 121384/30-1 FUGG, EV30/11-1, EV30/12-1, EV30/14-1 and SFB-1277(Projects A03). We gratefully acknowledge the Gauss Centre for Supercomputing e.V. (www.gauss-centre.eu) for funding this project by providing computing time on the GCS Supercomputer
SuperMUC at Leibniz Supercomputing Centre (www.lrz.de). This work was performed on the supercomputer ForHLR funded by the Ministry of Science, Research and the Arts Baden-Württemberg and by the Federal Ministry of Education and Research.

\section{Appendix\label{A1}}

\subsection{Experimental background}  \label{appendix_experiment}
\subsubsection{Experimental details}

A bulk Al crystal of (111) orientation was cleaned by cycles of Ar-ion sputtering and annealing to 700 K in UHV until clean and atomically flat surfaces were found with STM.  After STM inspection, impurities in form of
Fe atoms were deposited onto the sample while resting in the STM at cryogenic temperatures \cite{Crommie1993,Miyamachi13}. The low mobility at these temperatures prohibits thermal diffusion and the Fe atoms do not aggregate to larger clusters or islands.
At the base temperature of the STM of $\approx$ 25 mK \cite{Balashov18}, spectra of the voltage dependent tunneling current $I(U)$ were recorded with the feed-back loop of the STM disabled. $dI/dU$ curves were obtained by numerical derivative in order to avoid energy smearing when using a lock detection. STM tips were prepared from high purity W wire by chemical etching and cleaning in UHV. The tips were not superconducting as tested by taking tunneling spectra of a Au(111) surface at the base temperature.

\subsubsection{Gap enhancement with Fe-impurity}

\begin{figure}[t]
\includegraphics[width=0.9\columnwidth]{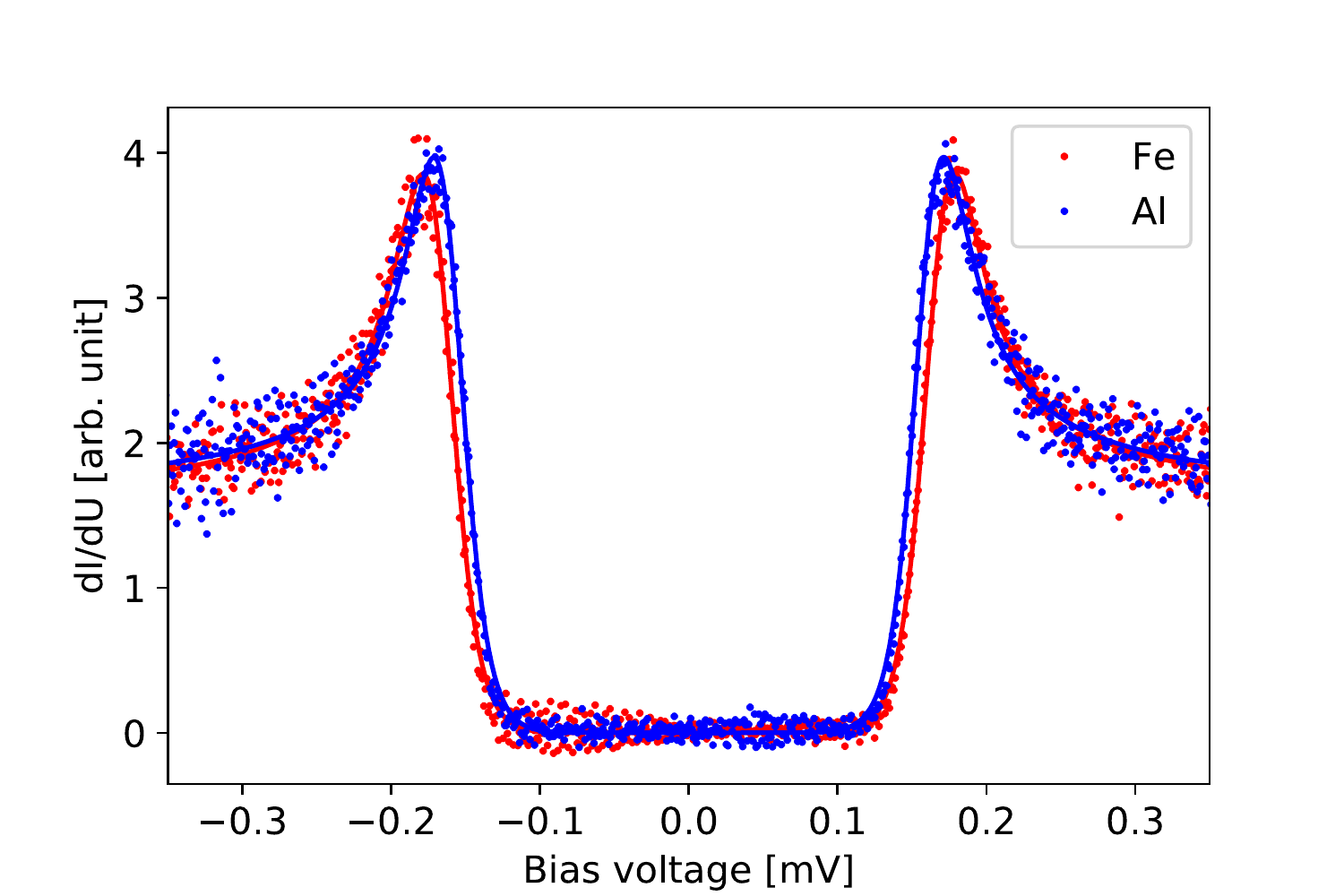} 
\caption{Differential conductance $dI/dU$ recorded at 25 mK on bare Al(111) (blue dots) and atop single Fe atom on Al(111) (red dots) together with fits (solid lines) to thermally smeared BCS density of states.  Blue and red solid lines display the fits to the experimental data points. Feed-back conditions: $U$= 400 $\mu$V, $I$= 500 pA.}
\label{Fig1} 
\end{figure}

Figure \ref{Fig1} shows the tunneling spectra recorded on a clean area $\approx$23 nm away from Fe atoms (blue dots) and atop an Fe impurity (red dots). Fe is the most common impurity in Al and the absence of a Kondo effect in high purity Al indicates that the magnetic moment of Fe in Al is absent from the beginning. To verify this, we drove Al to the normal phase by applying a magnetic field of 14 mT normal to the surface. While this field induces a transition to the normal state, it hardly is strong enough to eventually eliminate a possible Kondo resonance of the Fe spin. Nevertheless, we could not observe any Kondo peak in $dI/dU$ spectra (not shown) in agreement with the non-magnetic nature of Fe on Al(111). Thus with  the absence of a magnetic moment of Fe in Al, we also do not expect to observe Yu-Shiba-Rusinov bound states \cite{Yu1965,Shiba1968,Rusinov1968}, i.e. the Fe impurity does not break Cooper pairs by spin-flip scattering reducing the superconducting gap $\Delta$ or inducing in-gap states. Even more, $\Delta$ atop the Fe atoms appears larger than that on the bare Al(111) surface in the experiment (see Figure \ref{Fig1}). 

Fitting the $dI/dV$ spectra with thermally broadened BCS density of states \cite{Balashov18} allowed to quantify the increase from the bare Al gap of 159.6$\pm$0.1 $\mu$eV to 166.6$\pm$0.1 $\mu$eV, while the broadening of both curves due to the electronic temperature does not change significantly (112$\pm 1$ versus 115$\pm$1 mK). 
We conclude that, while Fe acts as a local scatterer for electrons and presumably also Cooper pairs on the Al(111) surface, it does not act as a spin scatterer. Following this result, any scatterer on the surface or near the surface should induce similar local variations of the $\Delta$ as evidenced for the buried defect in Fig. 1.

\subsection{Computational model and method\label{a2}} 
We study the Bogolibov-deGennes (BdG) model with an added Anderson impurity
\begin{equation}
\label{eq:Ham_friedel_2d}
\hat{H} = \hat{H}_\text{BdG} + \hat{H}_\text{imp},
\end{equation}
where
\begin{eqnarray}
\hat{H}_\text{BdG} &=&
- t\sum_{\langle i,j\rangle , \sigma} \hat{c}_{i,\sigma}^\dagger \hat{c}_{j,\sigma} - \sum_{i=1, \sigma}^{N_\text{lat}} \left(\frac{U}{2} n(\mathbf{r}_i)+\mu\right)  \hat{n}_{i,\sigma} \nonumber\\
&&-\sum_{i=1}^{N_\text{lat}} \Delta(\mathbf{r}_i) \hat{c}_{i,\uparrow} \hat{c}_{i,\downarrow} + \text{h.c.}.
\end{eqnarray}
The impurity is realized as an extra site external to the Hubbard lattice 
\begin{eqnarray}
\hat H_\text{imp} &=& -t \hat{c}_{I,\sigma}^\dagger \hat{c}_{1,\sigma} - \left(\frac{U}{2} n(\mathbf{r}_I)+\mu +V_\text{imp} \right)  \hat{n}_{I,\sigma} \nonumber\\
&&-\Delta(\mathbf{r}_I) \hat{c}_{I,\uparrow} \hat{c}_{I,\downarrow} + \text{h.c.}, 
\end{eqnarray}
with local occupation number $n(\mathbf{r}_i)= \sum_\sigma \langle \hat{n}_{i,\sigma} \rangle$ , pairing amplitude 
$\Delta(\mathbf{r}_i)=\langle \hat{c}^\dagger_{i,\downarrow} \hat{c}^\dagger_{i,\uparrow} \rangle$, $U>0$, number of lattice sites $N_\text{lat}$ and Anderson impurity of potential $V_\text{imp}=-0.06t$ at site $\br_I$. The impurity strength has been chosen to model the difference in work function of Fe and Al. All computations are conducted at $T{=}0$; 
the chemical potential $\mu$ is adjusted to fix the particle density $\sum_i \frac{n(\br_i)}{N_\text{lat}}=n$. In the case of the 2D system $\hat{H}_\text{BdG}$ is defined on a periodic square lattice of linear size $L=121 a$, with lattice constant $a$. In 3D the cubic lattice that $\hat{H}_\text{BdG}$ is defined on, is periodic in x- and y-axis, whereas on the z-axis we impose open boundary conditions. In 3D the linear system size is $L=22$. The Anderson impurity is located on one of the surfaces. The density $n(\br)$ and pairing amplitude $\Delta(\br)$ are computed self-consistently up to tolerance $\alpha=10^{-5}$ in Fig. \ref{Fig4} and $\alpha=10^{-6}$ in Fig. \ref{Fig6}. $\Delta(\br)$ will be given in units of the clean gap $\Delta_\textbf{BCS}$ without the impurity. 
To compute the response, we take the difference of the self-consistent potentials with and without impurity excluding the impurity site.
For a more in-depth description of the solution of the BdG system, we refer to Ref.\cite{stosiek20}. All results have been computed with a full-diagonalization solver. 

\bibliography{literature}

\end{document}